\newcommand{\be}{\begin{equation}}
\newcommand{\ee}{\end{equation}}
\newcommand{\bea}{\begin{eqnarray}}
\newcommand{\eea}{\end{eqnarray}}

\newcommand{\p}[1]{(\ref{#1})}
\documentstyle{article}
\def\a{\alpha}\def\b{\beta}

\begin{document}
\begin{flushright}
ECM-UB-97/32 \\
UG-9/97\\
hep-th/9711043\\
\end{flushright}

\vspace{3truecm}
\renewcommand{\thefootnote}{\fnsymbol{footnote}}
\begin{center}
{\large\bf M-brane intersections from worldvolume superalgebras}

\vspace{1cm}
Eric Bergshoeff$^1$, Joaquim Gomis $^2$ and 
Paul K. Townsend $^2$\ \footnote{On leave from DAMTP,
University of Cambridge, United Kingdom.}

\vspace{0.5cm}
${}^1${\it Institute for Theoretical Physics, Nijenborgh 4,\\
9747 AG Groningen, The Netherlands}

\vspace{0.5cm}
${}^2$ Departament ECM, {\it Facultat de F{\'\i}sica,
Universitat de Barcelona and Institut de Fisica d'Altes Energies,\\
Diagonal 647, E-08028 Barcelona, Spain}

\vspace{1.cm}
{\bf Abstract}
\end{center}
It is known that the M-branes of M-theory correspond to p-form charges in the
D=11 spacetime supersymmetry algebra. Here we show that their intersections
are encoded in the p-form charges of their {\sl worldvolume} supersymmetry
algebras. Triple intersections are encoded in double intersection worldvolume
algebras with eight supercharges.

\renewcommand{\thefootnote}{\arabic{footnote}}
\setcounter{footnote}0
\newpage
\section{Introduction}

The 11-dimensional (D=11) M-theory is currently the leading candidate for a
unified theory of particle interactions, despite the fact that a complete
understanding of its microscopic degrees of freedom (especially those relevant 
to compactifications to four spacetime dimensions) remains elusive.
Fortunately, many of the implications of M-theory can be deduced
from semi-classical considerations. At this level, M-theory appears to be a
theory of one `basic', or 1/2-supersymmetric, `object' which can appear in
various dual forms. These can be found as solutions of the D=11 
supergravity field equations. Alternatively, the possibilities may be
investigated more abstractly via the `M-theory superalgebra' \cite{democracy};
this is a modification of the standard  D=11 supersymmetry algebra to
incorporate 2-form and 5-form central charges \cite{vans}. The space components
of these charges are associated \cite{azc,dmitri} with the M-2-brane and
M-5-brane. The time components of the 2-form charge may be associated
\cite{dmitri,hull} with a boundary of the D=11 spacetime, as in the Ho{\v
r}ava-Witten (HW) construction of the heterotic string \cite{HW}. For
terminological convenience we refer to this as the M-9-brane;
it can in fact be viewed as a decompactification limit of eight D-8-branes
coincident with an orientifold 8-plane in Type IA superstring theory
\cite{polwit}. The time component of the 5-form charge is associated with the
Kaluza-Klein (KK) monopole solution of D=11 supergravity, 
alias the D-6-brane of
IIA supergravity \cite{PKT}. We shall refer to this as the M-KK-monopole.
Finally, the space components of the 11-momentum are associated with the
massless quanta of D=11 supergravity or, at the classical level, with the
`M-wave' solution of D=11 supergravity \cite{Mwave}. This is the only 1/2
supersymmetric configuration permitted by the standard D=11 supersymmetry
algebra, hence the necessity of considering the full M-theory superalgebra.

To summarize, there are five `basic', 1/2-supersymmetric, constituents of
M-theory. These are
\be\label{onea}
M-Wave,\qquad M-2-brane,\qquad M-5-brane,\qquad M-KK-monopole,\qquad M-9-brane.
\ee
The M-2-brane and M-5-brane are often referred to collectively as `M-branes'; 
they are in fact the only {\it bone fide} branes. All the same, we shall find
it convenient to refer collectively to all five `basic' constituents as
`M-branes'. M-theory configurations with less than 1/2 supersymmetry may now be
found as `intersecting' M-branes \cite{paptown}. 
The simplest cases are those in
which two M-branes intersect to yield a configuration 
with 1/4 supersymmetry (of
the M-theory vacuum). From the point of view of each of the two constituent 
branes, the intersection appears as a 1/2 supersymmetric object in its
worldvolume. Thus, one way to study 1/4-supersymmetric intersections is by
looking for 1/2 supersymmetric solutions of the {\sl worldvolume} field
equations of the various M-branes. For example, it is known that two M-5-branes
may intersect on a 3-brane, preserving 1/4 supersymmetry \cite{paptown,tsey},  
and that an M-2-brane may end on an M-5-brane \cite{strom,town}, again 
preserving 1/4 supersymmetry (the boundary may be considered as a string
intersection for present purposes). From the perspective of the M-5-brane (or
one of them if there is more than one) these
intersections should appear as 1/2 supersymmetric 3-brane and string solutions
of its effective worldvolume field equations. Both solutions have recently been
found
\cite{HLW1,HLW2}, following the discovery of analogous solutions of the
worldvolume field equations of D-branes
\cite{CM,GWG}. It is remarkable that these solutions, while ostensibly just
configurations of a particular worldvolume field theory, actually provide their
own {\it spacetime}  interpretation as intersecting branes; this comes about
because the worldvolume scalars determine the spacetime embedding.  

Here we take this a step further by showing that the {\sl spacetime}
interpretation of a worldvolume p-brane preserving 1/2 supersymmetry is encoded
in the {\sl worldvolume} supersymmetry algebra. Indeed, {\sl all} 1/4
supersymmetric M-brane intersections may be deduced this way. The first point
to appreciate is that worldvolume p-branes are associated with p-forms in the
worldvolume supersymmetry algebra. In itself, this is not surprising given
our experience with the spacetime supersymmetry algebra. For example, it was
pointed out in \cite{HLW2} that the maximal central extension of the D=6 (2,0)
worldvolume supertranslation algebra of the M-5-brane has the anticommutator
\be\label{oneb}
\{Q_\a^I,Q_\b^J\} = \Omega^{IJ} P_{[\a\b]} + Y^{[IJ]}_{[\a\b]} +
Z^{(IJ)}_{(\a\b)}\, , \qquad (\Omega_{IJ}Y^{[IJ]} = 0)\, ,
\ee
where $\a,\b =1,\dots,4$ is an index of $SU^*(4)\cong Spin(5,1)$ and
$I=1,\dots,4$ is an index of $Sp(2)$, with $\Omega^{IJ}$ being its
invariant antisymmetric tensor. The $Y$-charge is a worldvolume 1-form and the
$Z$-charge a worldvolume self-dual 3-form. The presence of such charges in the
worldvolume supersymmetry algebra is implied by the existence of the 1/2
supersymmetric string and 3-brane solutions of the M-5-brane worldvolume field
equations. There is still a puzzle to be resolved, however. 
There are a total of
five  1-form charges and ten 3-form charges 
(in the irreducible {\bf 5} and {\bf
10} representations of $Sp(2)$), which exceeds 
the number of distinct worldvolume
strings and 3-branes. In the spacetime supersymmetry algebra, every p-form
charge corresponds to a distinct p-brane. 
This is clearly not so for worldvolume
algebras. The resolution of this puzzle is simply that the $Sp(2)$
representations provide the information needed to reconstruct 
the {\it spacetime}
interpretation of the branes within the M-5-brane. For example, identifying
$sp(2)\cong Spin(5)$ as the double cover of the $SO(5)$ rotation group
transverse to the M-5-brane worldvolume in the D=11 spacetime, we see that
the 1-forms $Y$ define a vector in this transverse 5-space. This information
serves to identify the 1-brane in the worldvolume as the boundary of an 
M-2-brane in the M-5-brane. Note, however, that the {\bf 5} representation of
$Sp(2)$ could equally well have been interpreted as defining  a 4-form in the
transverse 5-space, so the 1-brane has another spacetime interpretation as the
intersection of the M-5-brane with another M-5-brane, which is again an M-brane
intersection known to preserve 1/4 supersymmetry \cite{GKT}.  

So far we have implicitly considered only the space components of
$Y$. As pointed out in \cite{HLW2}, the time components can be viewed as the
space components of a worldvolume dual 5-form and are therefore charges for a
space-filling 5-brane in the M-5-brane. We interpret this to mean that the
given M-5-brane is actually inside another brane, so that its `intersection'
with it coincides with the M-5-brane. If the {\bf 5} representation is taken to
define a 1-form in transverse 5-space we might then conclude that the M-5-brane
is inside a 6-brane. There is no M-theory 6-brane as such but there is the
M-KK-monopole. Adopting this interpretation we conclude, correctly \cite{TC},
that an M-5-brane `inside' an M-KK-monopole preserves 1/4 supersymmetry. If
instead we take the {\bf 5} representation to define a transverse 4-form then
we conclude, again correctly \cite{deRoo}, that an M-5-brane inside an
M-9-brane preserves 1/4 supersymmetry; in fact, this is just the heterotic
5-brane in the HW-formulation of the heterotic string theory.

There is still much to learn from the algebra \p{oneb}. 
For example, we have not
yet considered the $Z$ charges in the {\bf 10} representation of $Sp(2)$.
Suffice it to say that the complete analysis precisely duplicates the known
classification of 1/4 supersymmetric intersections involving an M-5-brane.
There remains a question of consistency. To take an example, we have just
deduced from the M-5-brane worldvolume superalgebra that it can have a
1/4-supersymmetric intersection with an M-2-brane. We should be able to see
this from the M-2-brane's worldvolume algebra. This is (after gauge fixing)
a D=3 field theory with N=8 supersymmetry. Allowing for all possible p-form
central charges in the N=8 D=3 supertranslation algebra we arrive at the
anticommutator
\be\label{onec}
\{Q_\a^i,Q_\b^j\} = \delta^{ij} P_{(\a\b)} + Z_{(\a\b)}^{(ij)} +
\varepsilon_{\a\b} Z^{[ij]}\, , \qquad (\delta_{ij}Z^{(ij)}=0)\, ,
\ee
where $Q^i$ are the eight D=3 Majorana spinor supercharges ($i=1,\dots,8$), and
$P$ is the 3-momentum\footnote{This, at least, is the obvious interpretation of
this `charge'. We shall see later that it has another interpretation.}. This
supersymmetry algebra has an
$SO(8)$ automorphism group, which we interpret as the rotation group in the
transverse 8-space. The supercharges do not transform as an 8-vector, however,
but as a chiral $SO(8)$ spinor. The 0-form and 1-form central charges are
therefore
\be\label{oned} 
Z(0-form): {\bf 28}\, , \qquad Z(1-form): {\bf 35}^+ \, ,
\ee
where the ${\bf 28}$ is either a 2-form or 6-form of the transverse 8-space and
the ${\bf 35}^+$ is a self-dual transverse 4-form. Consider first the
worldvolume 0-forms, corresponding to a 0-brane on the D=3 worldvolume. 
Choosing
the 2-form interpretation of the ${\bf 28}$ representation we see that the
0-brane is the result of an intersection with another M-2-brane, a known 1/4
supersymmetric M-brane intersection \cite{paptown}. If we instead
choose the 6-form interpretation of the ${\bf 28}$ then the worldvolume 0-brane
acquires the interpretation as the intersection of the M-2-brane with a
KK-monopole \cite{bergs}. 
Consider now the worldvolume 1-forms, corresponding to
a 1-brane on the D=3 worldvolume. 
This 1-brane is determined by a 4-plane in the
transverse space and is therefore the intersection of an M-5-brane with the
M-2-brane. We therefore recover from the M-2-brane worldvolume superalgebra the
intersection with the M-5-brane that we previously deduced from the latter's
worldvolume superalgebra. Again, there is much more to be learnt from the
algebra \p{onec}, but we postpone the complete analysis. 

The remaining M-brane superalgebras can be dealt with in the same way. The
final result is that each 1/4-supersymmetric intersection of two M-branes
occurs precisely once in each of the lists of intersections derived from the
worldvolume supersymmetry algebras of its constituents. If the same analysis is
applied to D-branes the results are not quite so nice. It remains true
that every intersecting brane configuration preserving 1/4 supersymmetry
corresponds to some spacetime interpretation of a p-form charge in the
worldvolume superalgebra of either participating brane. Conversely, every
p-form in a worldvolume superalgebra has a spacetime interpretation as a
1/4-supersymmetric configuration of intersecting branes, but it is no
longer always the case that {\it every} such interpretation corresponds to a
1/4-supersymmetric configuration; one has to impose additional {\it ad hoc}
interpretational rules. The reason for this can be traced to some peculiarities
of supersymmetry algebras with 16 supercharges. The complete list of such
algebras is as follows
$$
\begin{array}{cccccc}
 & D=1 & N=16 & so(9) & {\bf 16} & D-0\\
\ast & D=2 & (16,0) & so(9) & {\bf 16} & M-W \\
 & D=2 & (16,0) & so(8) & {\bf 8}_s \oplus {\bf 8}_s & IIB-W \\
 & D=2 & (8,8) & so(8) & {\bf 8}_s \oplus {\bf 8}_c  & IIB-1\\
 & D=2 & (8,8) & so(8) & {\bf 8}_s \oplus {\bf 8}_s  & IIA-1\\
\ast & D=3 & N=8   &  so(8) & {\bf 8}_s & M-2 \\
 & D=4 & N=4   & so(6) & {\bf 4} \oplus \bar{\bf 4} & D-3 \\
 & D=5 & N=2 & so(5) & {\bf 4} & D-4 \\
\ast & D=6 & (2,0) & so(5) & {\bf 4} & M-5 \\
 & D=6 & (1,1) & so(4) & {\bf 4} &IIB-5 \\
\ast & D=7 & N=1 & so(3) & {\bf 2} & D-6,\, M-KK \\
 & D=8 & N=1 & so(2) & {\bf 2} & D-7 \\
 & D=9 & N=1 & Z_2 & - & D-8 \\
\ast & D=10 & (1,0) & - & -&  D-9,\, M-9 
 \end{array}
$$
The value of N is given in the third column, with the convention that minimal
supersymmetry is always N=1; for D=2,6,10, the number of chiral and
antichiral spinors must be specified separately. The automorphism, or
`R-symmetry', algebra is given in the fourth column; the group is generally the
spinorial covering group and the representation to which 
the supercharges belong
is given in the fifth column. For D=1,2 we have not given the largest possible
automorphism algebra. For example, the (8,8) algebras actually have a $Spin(8)$
group acting independently on spinors of each chirality, but in the
string-theory realization only the diagonal $Spin(8)$ survives. Similarly, in
D=1 the maximal automorphism group is $SO(16)$ but only $SO(9)$ is realized by
0-branes. This is because the supercharges must be spinors of the R-symmetry
group as well as of the worldvolume Lorentz group, and $SO(9)$ is the largest
subgroup of $SO(16)$ with a 16-dimensional spinor representation. In the last
column we give a brane field theory that realizes the supersymmetry algebra.
Some branes do not appear in the list because they  realize only a proper
subgroup of the automorphism algebra as a consequence of being the dimensional
reduction of another brane with the same algebra. For example, the IIA-wave is
essentially the same as the M-wave but with the $SO(9)$ symmetry reduced to
$SO(8)$. 

If one associates the $so(n)$ R-symmetry algebra of each of the above
supersymmetry algebras to an $SO(n)$ rotation group in a space transverse to a
(p+1)-dimensional worldvolume then one arrives at an interpretation in which
the p-brane is embedded in a larger spacetime of total dimension $D=(n+p+1)$.
This total dimension is in each case either D=10 or D=11. This fact might well
be considered remarkable to someone studying these algebras as a mathematical
excercise, but it is of course expected 
from their interpretation as worldvolume
superalgebras of D=10 or D=11 branes. Note that a single transverse dimension
would be associated with a trivial 
$SO(1)$ rotation group so that in those cases
for which the R-symmetry algebra is trivial there is an ambiguity in the
dimension of the transverse space: it could be zero or one. For example, if we
suppose that the transverse space associated to the D=10 (1,0) algebra has
dimension one we may interpret the algebra as that of the M-9-brane. If instead
we choose the transverse space to have dimension zero we may interpret the
algebra as that of the D-9-brane, 
which is effectively the D=10 spacetime of the
Type I superstring. The D=9 N=1 supersymmetry algebra is another rather special
case. Its $Z_2$ automorphism group is the vestige of a rotation group in a
1-dimensional transverse space, so the D-8-brane acquires a natural D=10
interpretation but, as usual, its 11-dimensional interpretation is obscure. 

An inspection of the above supersymmetry algebras reveals that most are
dimensional reductions of some other one. The only exceptions are the five
supersymmetry algebras corresponding to the worldvolume field theories of the
five M-branes. These have been marked with an asterisk in the first
column. The M-wave, M-5-brane and M-9-brane algebras are manifestly not
reductions of any other algebra in the list. Dimensional reduction of the
D-3-brane algebra yields an N=8 D=3 algebra but its automorphism algebra is
$so(7)$ rather than $so(8)$. This is of course the algebra of the
D-2-brane, which we did not include because it is just the reduction of the
M-2-brane. Thus, the M-2-brane algebra is also not the reduction of any other
one. This leaves the algebra associated to the M-KK-monopole. 
Strictly speaking,
this is an exception to the rule because it is also the 
D-6-brane algebra, which
is  the reduction of the D-7-brane algebra. 
However, this algebra shares the nice
features of the other M-brane algebras, and when one takes into account the
hidden dimension implicit in the KK-monopole metric, there is a natural D=11
interpretation. In any case, its inclusion among the `exceptional' cases is
forced by the intersecting brane interpretation of the p-form charges in the
other M-brane algebras. 
We learn from all this that when determining the possible
1/4-supersymmetric brane intersections from the worldvolume supersymmetry
algebras of participating branes we need only consider the five M-branes. Thus,
the worldvolume superalgebras of D=10 and D=11 branes proclaim the `supremacy'
of M-theory.

Having determined all 1/4-supersymmetric intersections of M-branes from the
supersymmetry algebras with 16 supercharges one is led to wonder whether
intersections preserving less supersymmetry are encoded in the supersymetry 
algebras with fewer supersymmetry charges. The answer appears to be a
qualified `yes', but we shall not attempt a 
systematic analysis here. Rather, we
shall present a few illustrative cases in which triple intersections with 1/8
supersymmetry can be deduced from the p-form charges in the superalgebra with 8
supercharges corresponding to 1/4-supersymmetric pairwise brane intersections.
One case of interest is Type I brane intersections viewed as triple
intersections in which the `third brane' is a D-9-brane. Another class of cases
can be deduced from the worldvolume algebra of the 3-brane in the M-theory
5-brane. Analysis of the p-forms in this algebra shows, in particular, that 
the 3-brane in the M-5-brane is a D-brane for the self-dual string. We shall
provide both D=10 IIB and D=11 M-theory explanations of this fact.

\section{M-brane intersections}

\subsection{M-wave}

The M-wave algebra is
\be\label{mwa}
\{Q_-^i, Q_-^j\} =\delta^i_j P_= + Z_=^{(ij)}\, , 
\qquad (\delta_{ij}Z_=^{(ij)}=0)\, ,
\ee
where the supercharges $Q_-^i\ (i=1,\cdots ,16)$ 
are in the {\bf 16} spinor representation of
$SO(9)$. This has the obvious reduction to the D=10 D-0-brane algebra
\be\label{mwb}
\{Q^i, Q^j\} =\delta^i_j H + Z^{(ij)}({\bf 126}\oplus {\bf 9})\, .
\ee
As indicated, the 0-form central charges in this algebra fall into the {\bf
126} and {\bf 9} representations of $SO(9)$, as do the D=2 self-dual vector
Z-charges in the M-wave algebra. Intersections of M-waves preserving 1/4
supersymmetry have been classified in \cite{bergs}, but since these are less
familiar than those involving D-0-branes it is helpful to consider the latter
first. The results for the M-wave are then obtained by re-interpreting the
D-0-brane results in D=11. 

The {\bf 126} of $SO(9)$ is either a 4-form or a 5-form in the transverse
9-space. This corresponds to a D-0-brane in a D-4-brane or a KK-monopole,
respectively. The {\bf 9} of $SO(9)$ is either a 1-form
or an 8-form in the transverse 9-space. This corresponds to a D-0-brane
in a fundamental string or a D-8-brane, respectively. 
The D=11 interpretation of the same configurations is, in the notation
of \cite{bergs},
\be\label{mwc}
(1|W,M5),\qquad (1|W,KK),\qquad (1|W,M2),\qquad (1|W,M9)\, .
\ee
This is the complete list of 1/4-supersymmetric intersections of M-waves.

\subsection{M-2-brane}

The M-2-brane algebra is given in \p{onec}. We considered earlier the spacetime
interpretation of the space components of the 1-form and 3-form charges. This
led to the following M2-brane intersections
\be\label{mtwoa}
(0|M2,M2),\qquad (0|M2,KK),\qquad (1|M2,M5)\, .
\ee
The time components of the 1-form charges are equivalent to space
components of the dual 2-forms and hence correspond to a `worldvolume 2-brane',
i.e. worldvolume filling 1/2-supersymmetric 2-branes. These cases are naturally
interpreted as `intersections' in which the original M-2-brane is inside some
larger brane. The latter is nominally a 6-brane since it is determined by an
additional 4-plane in the transverse space but, as earlier, this is to be
interpreted as a KK-monopole. We thus conclude that an M-2-brane inside a
KK-monopole preserves 1/4 supersymmetry, as indeed it does. This yields the
intersection 
\be\label{mtwob}
(2|M2,KK)\, .
\ee
To find the remaining intersections of M-2-branes we must
take into account that the algebra \p{onec} admits 1/2 supersymmetric charge
configurations in which only the 3-vector $P$ is non-zero; it must
then be null. Since $P$ is also a singlet of $SO(8)$ we
may consider it to be either a 0-form in the transverse 8-space or an
8-form. In the former case $P$ has the obvious
interpretation as the null 3-momentum of an M-wave `inside' an M-2-brane, a
known 1/4-supersymmetric `intersection'. In the latter case $P$ represents
the intersection of a 9-dimensional object with the M-2-brane; it must be
interpreted as a charge associated to the boundary of an M-2-brane on an
M-9-brane. The fact that $P$ is null is an indication that the string
boundary is chiral, consistent with its HW-interpretation as the heterotic
string. We have now found the additional M2-brane intersections
\be\label{mtwoc}
(1|M2,W), \qquad (1|M2,M9)\, ,
\ee
and these complete the list of possible 1/4-supersymmetric intersections
involving M-2-branes.

\subsection{M-5-brane}

The M-5-brane worldvolume algebra is given in \p{oneb}. We have already
explained how the $Y$-charge leads to the intersections
\be\label{Mfivea}
(1|M5,M2),\qquad (1|M5,M5),\qquad (5|M5,KK),\qquad (5|M5,M9)\, .
\ee
Consider now the self-dual 3-form charge $Z$ carried by a 3-brane in the
M-5-brane. This charge is in the {\bf 10} representation of $Sp(2)$, which may 
be interpreted as either a 2-form or a 3-form in the transverse 5-space. In the
2-form interpretation, the 3-brane is the intersection with a second M-5-brane.
In the 3-form interpretation, it is the intersection with an M-KK-monopole. We
thus deduce the intersections
\be\label{Mfiveb}
(3|M5,M5),\qquad (3|M5,KK)\, .
\ee
Finally, we must consider the $SU^*(4)\cong Spin(5)$ 5-vector $P$. If $P$ is
assumed  to be a 0-form in the transverse space then $P$ has the obvious
interpretation as the null 5-momentum of an M-wave in the M-5-brane. 
However, we
could also suppose $P$ to be a transverse 5-form, which would indicate a
6-dimensional object. This must be interpreted as an M-KK-monopole intersecting
the M-5-brane on a chiral string. We thus deduce the
intersections
\be\label{Mfivec}
(1|M5,W), \qquad (1|M5,KK)\, .
\ee
Note that if the non-compact hyper-Kahler 4-metric of the KK-monopole were
replaced by a compact hyper-Kahler metric on $K_3$ then the `intersection
$(1|M5,KK)$ would become an M-5-brane wrapped on $K_3$, alias the heterotic
string \cite{pktb}. This goes some way to explaining why the string
intersection in $(1|M5,KK)$ is chiral. In any case, we have now found the
complete set of possible M-5-brane intersections preserving 1/4 supersymmetry.

\subsection{M-KK-monopole}

The M-KK-monopole algebra is 
\bea\label{KKa}
\{Q_\a^i,Q_\b^j\} &=& \varepsilon^{ij}(C\Gamma_\mu)_{\a\b}P^\mu + 
C_{\a\b} Z^{(ij)} \\
&& + \varepsilon^{ij}(C\Gamma^{\mu\nu})_{\a\b} Z_{\mu\nu} +
(C\Gamma^{\mu\nu\rho})_{\a\b} Z_{\mu\nu\rho}^{(ij)}\, ,\nonumber
\eea
where $i=1,2$ is an $Sp(1)\cong SU(2)$ index. We therefore have the following
p-brane charges with their $SU(2)$ representions
\be\label{KKb}
Z(p=0): {\bf 3} \qquad Z(p=2) : {\bf 1} \qquad Z(p=3): {\bf 3} \qquad
Z(p=4): {\bf 3} \qquad Z(p=5) : {\bf 1}
\ee
The p=4,5 cases come from the time components of the 3-form and 2-form
charges. The algebra \p{KKa} is also the algebra of the D-6-brane and it will 
be convenient to begin with an analysis of its implications for D-6-brane
intersections. In this case the transverse space is three-dimensional, so an
$SU(2)$ singlet can be interpreted as either a 0-form or as a 3-form in
the transverse 3-space. Similarly, an $SU(2)$ triplet can be interpreted as
either a 1-form or as a 2-form in the transverse space. Each possibility yields
a distinct configuration of intersecting branes. The details are much as
before and the result is that one finds all 1/4-supersymmetric intersections of
D-6-branes for which the intersection is a p-brane with p=0,2,3,4,5. There
remain two p=1 intersections not yet accounted for. One is a IIA-wave in a
D-6-brane which is obviously associated with $P$, interpreted as
the null 7-momentum of the wave. The other is found by taking the space
components of the null 7-vector $P$ to be a 3-form in transverse 3-space. In
this case $P$ can be interpreted as the charge of a chiral string formed by the
intersection of a D-4-brane with the D-6-brane. This can be represented by the
array
$$
\begin{array}{lcccccccccc}
D6: & 1 & 2 & 3 & 4 & 5 & 6 & - & - & - \\
D4: & 1 & - & - & - & - & - & 7 & 8 & 9 
\end{array}
$$
Note that there are no dimensions in which the D-4-brane could be separated
from the D-6-brane, so that a chiral theory on the intersection is
indeed possible (the 11th dimension does not invalidate this argument because
the M-KK-monopole is not strictly a brane with a localized worldvolume). 
In fact,
this configuration is dual (at least formally) to a D-string in a D-9-brane,
alias the (chiral) type I superstring. 

Let us now return to the M-KK-monopole. The transverse space is now
4-dimensional, despite the $Spin(3)$ R-symmetry. This means that the transverse
dual to, say, a transverse 1-form is now a 3-form, whereas it was previously a
2-form. However, what was previously a 1-form might now really be a
two-form, in which case the transverse dual would still be a 2-form. In other
words, in associating forms of two possible ranks to each worldvolume p-form in
the algebra, one of the two ranks will now increase by one, the other one being
unaffected. There is therefore an additional choice that must now be made,
which amounts to a choice of where to put the extra dimension when
re-interpreting the previous D-6-brane results in D=11. The choice is not
arbitrary, however, because it is determined by the possible M-branes. Only one
choice has a D=11 interpretation because there is only one way to
interpret each D-6-brane intersection as a KK-monopole intersection. In the
case of 0-brane intersections, the two D-6-brane cases lift to the
intersections
\be\label{KKc}
(0|M2,KK)\, , \qquad (0|{\underline M2},KK)\, ,
\ee
where the underlining in the second case indicates that the compact isometry
direction of the KK-monopole is also an isometry of the M2-brane; in other
words, the M-2-brane is wrapped on the KK circle. The two D-6-brane cases with
2-brane intersections lift to
\be\label{KKd}
(2|M2,KK)\, , \qquad (0|KK,KK)\, .
\ee
The two 3-brane intersections are
\be\label{KKe}
(3|M5,KK)\, , \qquad (3|{\underline M5}, KK)\, ,
\ee
with the same meaning to underlining as before. Similarly, the 4-brane
intersections are
\be\label{KKf}
(4|KK,KK)^a\, , \qquad (4|KK,KK)^b\, ,
\ee
where the notation is that of \cite{bergs}: the `a' case is the one in which
the KK circle is common to both KK `monopoles' (the arrays and supergravity
solutions may be found in \cite{bergs}). The 5-brane intersections are
\be\label{KKg}
(5|M5,KK)\, , \qquad (5|M9,KK)\, .
\ee
We thus recover from the M-KK algebra the M-5-brane intersection with an
M-KK-monopole deduced previously from the M-5 algebra. In addition we find
the KK intersection with an M-9-brane \cite{deRoo}; this can be interpreted as
a KK-monopole having an M-5-brane boundary on an M-9-brane.

Finally, the wave/chiral string intersections in D=11 are
\be\label{KKh}
(1|W,KK)\, , \qquad (1|M5,KK)\, ,
\ee
which we deduced earlier from the M-W and M-5 algebras.

\subsection{M-9-brane}

The M-9-brane worldvolume algebra is
\bea\label{Mninea}
\{Q_\a,Q_\b\} &=& (C{\cal P}^+\Gamma_\mu)_{\a\b}P^\mu + 
(C\Gamma^{\mu\nu\rho\sigma\lambda})_{\a\b} Z^+_{\mu\nu\rho\sigma\lambda}\, ,
\eea
where ${\cal P}^+$ is the chiral projection operator on spinors and $Z^+$ is a
self-dual 5-form. When this is interpreted as the D-9-brane algebra the
5-form charge is associated with a 5-brane in the D-9-brane, alias the Type I
D-5-brane. We shall consider this interpretation in more detail in the
following section. For its M-9-brane interpretation 
we must suppose that there is
a one-dimensional transverse space. The 5-form
$Z^+$ could represent either a 0-form or a 1-form in this transverse `1-space'.
In the 0-form case we have an M-5-brane inside an M-9-brane, 
alias the heterotic
5-brane. In the 1-form case we have what might appear to be a 5-brane
intersection of the M-9-brane by a 6-brane, 
but this should be interpreted as an
M-KK-monopole with a boundary on the M-9-brane. Thus, we confirm the cases
\be\label{Mnineb}
(5|M9,M5),\qquad (5|M9,KK),
\ee
found earlier from the M-5 and M-KK algebras. To find the remaining M-9-brane
intersections we note that the space components of the 10-momentum $P$ could be
either 0-forms or 1-forms in the transverse space. In the former case we have
the obvious interpretation as an M-wave in an M-9-brane. In the latter case the
object intersecting the M-9-brane is two-dimensional; it must be an M-2-brane
with boundary on the M-9-brane. 
The fact that $P$ is null is again an indication
of the chirality of the string boundary, which is in fact the heterotic string.
Thus we deduce the intersections
\be\label{Mninec}
(1|M9,W) ,\qquad (1|M9,M2)\, ,
\ee
which we also found previously from the M-W and M-2 algebras. We have now found
the complete set of M-9-brane intersections preserving 1/4-supersymmetry.

\section{Triple intersections}

We now turn to triple orthogonal intersections of M-branes or Type II
branes preserving 1/8 supersymmetry. The common-intersection worldvolume field
theory realizes a supersymmetry algebra with 8 supercharges. The list of such
algebras is as follows
$$
\begin{array}{cccccc}
 & D=1 & N=8 & so(8) & {\bf 8}_s & \\
\ast & D=2 & (8,0) & so(8) & {\bf 8}_s & IB-1 \\
 & D=2 & (4,4) & so(4) & {\bf 4} \oplus {\bf 4}  & \\
& D=3 & N=4   & so(4) & {\bf 4} & m-2\\
 & D=4 & N=2   & u(2) & {\bf 2}\oplus \bar {\bf 2}  &   \\
 & D=5 & N=1 & su(2) & {\bf 2} &   \\
\ast & D=6 & (1,0) & su(2) & {\bf 2} & 1B-5  
\end{array}
$$
The algebras marked with asterisks are those which are not the dimensional
reduction of any other. These are the worldvolume algebras of the Type 1 string
and 5-brane, which we call the IB-string and IB-5-brane (because of the IIB
origin of the Type 1 string). Double intersections of these 1B-branes,
preserving 1/4 of the supersymmetry of the Type 1 vacuum, can be considered as
special cases of 1/8-supersymmetric triple intersections of IIB branes in which
the `third brane' is the D-9-brane. We shall consider these cases first.

The Type 1 spacetime supersymmetry algebra is the same as the M-9-brane
worldvolume algebra considered above. As in that case, the null 10-vector $P$
does  `double duty' as a wave momentum and chiral string charge, while $Z^+$
does `double duty' as the charge of a 5-brane and a KK-monopole. 
In the past, the
necessity for charges to do `double duty' in the N=1 D=10 supersymmetry
algebra has always seemed a drawback of the approach but we have just seen that
it has a natural explanation from the M-9-brane perspective (as a result of the
0-form and 1-form interpretations in transverse space). In any case, we may now
attempt to determine the intersection of Type 1 branes by considering the
supersymmetry algebras of the 1B-string/1B-wave and the
1B-fivebrane/1B-KK-monopole. For simplicity we restrict the discussion to 
the 1B string and 5-brane. The 1B-string
worldsheet supertranslation algebra is the D=2 (8,0) algebra\ ($i=1,\dots,8$)
\be\label{threea}
\{Q_-^i,Q_-^j\} = \delta^{ij}P_= + Z_=^{(ij)}({\bf 35}^+)\, , \qquad
(\delta_{ij}Z_=^{(ij)}=0).
\ee 
As indicated, the 1-form Z-charges are in the ${\bf 35}^+$ representation of
$SO(8)$, which defines a 4-form in the transverse 8-space. We conclude,
correctly, that the 1B-string inside a 1B-5-brane preserves 1/4 of the
supersymmetry of the Type 1 D=10 Minkowski vacuum. We should be able to see
this intersection from the 5-brane's worldvolume supersymmetry algebra. 
This is 
the D=6 (1,0) algebra\ ($i=1,2; \a = 1,\cdots ,4$)
\be\label{threeb}
\{Q^i_\a,Q^j_\b\} = \varepsilon^{ij}P_{[\a\b]} + Z_{(\a\b)}^{(ij)}\, .
\ee
The Z-charge is an $SU(2)$-triplet of self-dual D=6 3-forms. This is the charge
of a 3-brane in the 5-brane. The $SU(2)\cong Spin(3)$ R-symmetry of the
algebra (\ref{threeb}) suggests a spacetime interpretation in D=9,
but the $SU(2)$ doublet of supercharges can equally well be interpreted as a
spinor of $SO(4)$, in which case an $SU(2)$ triplet is to be interpreted as a
self-dual 2-form in a transverse 4-space. Thus, the 3-brane is the intersection
of two 1B-5-branes. This possibility is to be expected from the IIB origin of
the 1B string. We have still to find the 1B-string in the 5-brane. Since the
string is chiral it should be associated with a null 6-vector. The only
possibility is $P$, which therefore does `double duty' as a string charge and
the 6-momentum of a 1B-wave in the 1B-5-brane. 

Another class of 1/8-supersymmetric triple intersections is obtained by
considering double intersections of branes confined to the core of a
KK-monopole. For example, consider a IIB-3-brane in a IIB-KK-monopole.
The IIB-3-brane is effectively confined to a D=6 spacetime. In fact, this is
one realization of the iib-3-brane of the `little' D=6 IIB string theory, alias
iib-string theory\footnote{We have found the discussion of `little' theories in
\cite{AH} useful in orienting our ideas. We refer to that paper for further
references.}. The worldvolume supersymmetry algebra of the iib-3-brane is the
N=2 D=4 algebra. The $\{Q,Q\}$ anticommutator, in Weyl spinor notation, is
\bea\label{threed}
\{Q_\a^i,Q_{\dot\b\; j}\} &=& \delta^i_j P_{\a\dot\b} + W_{\a\dot\b}^i{}_j
\qquad (W_{\a\dot\b}^i{}_i =0)\, ,\cr
\{Q_\a^i,Q_\b^j\} &= & \varepsilon_{\a\b}\varepsilon^{ij} Z + 
Y_{(\a\b)}^{(ij)}\, .
\eea
One sees from this that the iib-3-brane admits 1/2-supersymmetric p-branes for
p=0,1,2,3. The R-symmetry group is $U(2)$. We interpret the $U(1)\cong Spin(2)$
factor as the rotation group of the 2-space transverse 
to the worldvolume of the
D=6 iib-3-brane, and the $SU(2)$ factor as the rotational isometry group of the
KK-monopole implicit in the iib-theory's D=10 spacetime realization. The
iib-p-brane charges are in the following $U(2)$ representations
\be\label{oneg}
Z(p=0) : {\bf 1}_{+1} \qquad W(p=1): {\bf 3}_0 \qquad Y(p=2): {\bf
3}_{+1} \qquad W(p=3): {\bf 3}_0
\ee
where the subscript on the $SU(2)$ representations indicates the $U(1)$ charge 
(Z and Y are complex). The 0-branes must be interpreted as intersections, or
endpoints, of iib-strings on iib-3-branes. Thus iib-3-branes are d-branes for
iib-strings, a fact that is evident from their D=10 spacetime interpretation as
IIB-branes in a KK-monopole. The 1-branes are intersections with a IIB-3-brane
for which the two orthogonal directions span a 2-cycle in the 4-space of the
KK-monopole (this is determined by a 2-form in 3-space if the IIB brane is not
wrapped on the KK circle and by a 1-form in 3-space if it is wrapped on the KK
circle). The 2-branes can be interpreted as intersections with other
D=6 3-branes that are actually IIB-5-branes with some directions in the
4-space of the KK-monopole. The 3-brane can be interpreted as the original
iib-3-brane inside a IIB-5-brane for which two directions are in the KK
4-space. 

Another interpretation of the iib-string and iib-3-brane is as the string
and 3-brane in the M-5-brane, so the intersections of iib-branes just discussed
have implications for triple M-brane intersections. 
For example, we saw that two
iib-3-branes can intersect on a string. We conclude that the same must be true
of two 3-branes in an M-5-brane, as it is because it is 
implied by the following
triple intersection of M-5-branes preserving 1/8 supersymmetry
\cite{paptown}
$$
\begin{array}{lcccccccccc}
M-5: & 1 & 2 & 3 & 4 & 5 & - & - & - & - & - \\
M-5: & 1 & 2 & 3 & - & - & 6 & 7 & - & - & - \\
M-5: & 1 & - & - & 4 & 5 & 6 & 7 & - & - & -
\end{array}
$$
The 3-branes in M-5-branes can also intersect on a 2-brane because this is
implied by the alternative triple M-5-brane intersection
$$
\begin{array}{lcccccccccc}
M-2: & 1 & 2 & 3 & 4 & 5 & - & - & - & - & - \\
M-5: & 1 & 2 & 3 & - & - & 6 & 7 & - & - & - \\
M-5: & 1 & 2 & - & - & 5 & 6 & - & - & - & -
\end{array}
$$
The p=3 case of \p{oneg} is realized by the triple M-5-brane intersection 
$$
\begin{array}{lcccccccccc}
M-2: & 1 & 2 & 3 & 4 & 5 & - & - & - & - & - \\
M-5: & 1 & 2 & 3 & - & - & 6 & 7 & - & - & - \\
M-5: & 1 & 2 & 3 & - & - & - & - & 8 & 9 & -
\end{array}
$$
The 3-brane intersection of the first two M-5-branes lies inside the third
M-5-brane. 

This leaves only the p=0 case of \p{oneg} to be accounted for as a triple
M-brane intersection. We have already deduced that it must be an endpoint on
a 3-brane of the iib-string, now in its guise as the self-dual string in the
M-5-brane. The fact that the iib string can end on the iib 3-brane can now be
seen from the triple M-brane intersection array
$$
\begin{array}{lcccccccccc}
M-5: & 1 & 2 & 3 & 4 & 5 & - & - & - & - & - \\
M-5: & 1 & 2 & 3 & - & - & 6 & 7 & - & - & - \\
M-2: & - & - & - & - & 5 & 6 & - & - & - & -
\end{array}
$$
This can be interpreted as the intersection of two M-5-branes on a 3-brane with
the 3-brane intersecting the string boundary of an M-2-brane. The membrane
boundary is continuous but switches from one M-5-brane to the other at the
intersection point. From the perspective of one of the M-5-branes the string
boundary appears to end on a 3-brane.

\section{Discussion}

We have seen that 1/4 supersymmetric intersections of M-branes, and Type II
superstring branes, are encoded in the supersymmetry algebras with 16
supercharges interpreted as worldvolume supersymmetry algebras. The
interpretation of the M-brane algebras is particularly natural. We have also
shown how some algebras with 8 supercharges encode triple intersections, but
the interpretation is less natural in the sense that there are various
ambiguities that arise which must be resolved by additional information. This
is probably inevitable because a 1/8 supersymmetric M-brane configuration, to
take an example, could be a non-orthogonal double intersection rather than an
orthogonal triple intersection. Thus, it is likely that the algebras with
less than 16 supercharges have several inequivalent interpretations as
intersection worldvolume algebras. However, we suspect that there
exists some natural iteration procedure that will at least allow a 
classification
of all orthogonal multiple intersections from the supersymmetry algebras with
16,8,4,2,1 supercharges. In any case, it is now clear that the various
supersymmetry algebras in various dimensions encode an enormous amount of
information about partially supersymmetric configurations 
of branes in M-theory.

\vskip 0.5cm
\noindent
{\bf Acknowledgements}: We thank George Papadopoulos for helpful discussions.
PKT gratefully acknowledges the support of the Iberdrola 
{\it Profesor Visitante} program. EB thanks the members of the Faculty of
Physics at the University of Barcelona for hospitality. This work has been
partially supported by AEN95-0590 (CICYT), GRQ93-1047 (GIRIT) and by the
Commission of European Communities CHRX93-0362 (04).


\bigskip
\begin{thebibliography}{99}

\bibitem{democracy}
P.K. Townsend, {\it p-brane democracy}, in {\it Particles, Strings and
Cosmology}, eds. J. Bagger, G. Domokos, A. Falk and S. Kovesi-Domokos (World
Scientific 1996), hep-th/9507048.

\bibitem{vans}
J.W. van Holten and A. Van Proeyen, {\sl J. Phys. A: Math Gen.} {\bf 15}
(1982) 3763.

\bibitem{azc}
J. A. de Azcarraga, J. P. Gauntlett, J. M. Izquierdo and P. K. Townsend,
{\sl Phys. Rev. Lett.} {\bf 63} (1989) 2443.

\bibitem{dmitri}
D. Sorokin and P.K. Townsend, {\sl M-theory superalgebra from the M-5-brane}, 
hep-th/9708003.

\bibitem{hull}
C.M. Hull, {\sl Gravitational duality, branes and charges}, hep-th/9705162.

\bibitem{HW}
P. Ho{\v r}ava and E. Witten, {\sl Nucl. Phys.} {\bf B460} (1996) 506; 
{\it idem} {\bf B475} (1996) 94.

\bibitem{polwit}
J. Polchinski and E. Witten, {\sl Nucl. Phys.} {\bf B460} (1996) 525.

\bibitem{PKT}
P.K. Townsend, {\sl Phys. Lett.} {\bf 350B}, (1995) 184.

\bibitem{Mwave}
C.M. Hull, {\sl Phys. Lett.} {\bf 139B} (1984) 39.

\bibitem{paptown}
G. Papadopoulos and P.K. Townsend, {\sl Phys. Lett.} {\bf 380B} (1996) 273.

\bibitem{tsey}
A. Tseytlin, {\sl Nucl. Phys.} {\bf B475} (1996) 149.

\bibitem{strom}
A. Strominger, {\sl Phys. Lett.} {\bf 383B} (1996) 44.

\bibitem{town}
P.K. Townsend, {\sl Phys. Lett.} {\bf 373B} (1996) 68.

\bibitem{HLW1}
P.S. Howe, N.D. Lambert and P.C. West, {\it The self-dual string soliton},
hep-th/9709014.

\bibitem{HLW2}
P.S. Howe, N.D. Lambert and P.C. West, {\it The threebrane soliton of the
M-fivebrane}, hep-th/9710033.

\bibitem{CM}
C.G. Callan and J.M. Maldacena, {\it Brane dynamics from the Born-Infeld
action}, hep-th/9708147.

\bibitem{GWG}
G.W. Gibbons, {\it Born-Infeld particles and Dirichlet p-branes},
hep-th/9709027.

\bibitem{GKT}
J.P. Gauntlett, D. Kastor and J. Traschen, 
{\sl Nucl. Phys.} {\bf B478} (1996) 544.

\bibitem{TC} A.A.Tseytlin, {\sl Nucl. Phys.} {\bf B487} (1997) 141;
             M. Costa, {\sl Nucl. Phys.} {\bf B490} (1997) 202.

\bibitem{deRoo}
M. de Roo, {\sl Intersecting branes and supersymmetry}, hep-th/9703124.

\bibitem{bergs}
E. Bergshoeff, M. de Roo, E. Eyras, B. Janssen and J.P. van der Schaar, {\sl 
Class. Quantum Grav.} {\bf 14} (1997) 2757.

\bibitem{AH}
R. Argurio and L. Houart, {\it Little Theories in Six and Seven Dimensions},
hep-th/9710027.

\bibitem{pktb}
P.K. Townsend, {\sl Phys. Lett.} {\bf 354B} (1995) 247.



\end{thebibliography}
\end{document}